\documentclass[twocolumn,nofootinbib,floatfix,amsmath,amssymb,showpacs,superscriptaddress]{revtex4}

\usepackage{graphicx}
\usepackage{dcolumn}
\usepackage{bm}
\usepackage{color}

\begin{document}

\title{Axial Charges of Octet Baryons in Two-flavor Lattice QCD}

\author{G\"{u}ray Erkol}
\affiliation{Laboratory for Fundamental Research, Ozyegin University, Kusbakisi Caddesi No:2 Altunizade, Uskudar Istanbul 34662 Turkey}
\author{Makoto Oka}%
\affiliation{Department of Physics, H-27, Tokyo Institute of Technology, Meguro, Tokyo 152-8551 Japan}
\author{Toru T. Takahashi}
\affiliation{Yukawa Institute for Theoretical Physics, Kyoto University,
Sakyo, Kyoto 606-8502, Japan}

\date{\today}

\begin{abstract}
We evaluate the strangeness-conserving $N N$, $\Sigma\Sigma$, $\Xi\Xi$, $\Lambda\Sigma$ and the strangeness-changing $\Lambda N$, $\Sigma N $, $\Lambda\Xi$, $\Sigma\Xi$ axial charges in lattice QCD with two flavors of dynamical quarks and extend our previous work on pseudoscalar-meson--octet-baryon coupling constants so as to include $\pi\Xi\Xi$, $K\Lambda\Xi$ and $K\Sigma\Xi$ coupling constants. We find that the axial charges have rather weak quark mass dependence and the breaking in SU(3)-flavor symmetry is small at each quark-mass point we consider.

\end{abstract}
\maketitle

\section{Introduction}
Hyperon axial charges are significant parameters for low-energy effective description of baryon sector as they enter in the loop graphs of chiral perturbation theory. While the nucleon axial charge can be precisely determined from nuclear $\beta$-decay (the modern value is $g_{A,NN}=1.2694(28)$~\cite{Amsler:2008zzb}), we do not have enough information about hyperon axial charges from experiment. The theoretical estimates from chiral perturbation theory~\cite{Savage:1996zd, Jiang:2008aqa, Jiang:2009sf}, large N$_c$ limit~\cite{FloresMendieta:1998ii} of QCD and QCD sum rules~\cite{Chiu:1985ka} exist but the results from these approaches are rather imprecise. The lattice calculations of the axial charge of the nucleon have reached a mature level~\cite{Edwards:2005ym, Khan:2006de, Yamazaki:2008py} however only recently there have been attempts to extract the hyperon axial charges using lattice QCD~\cite{Lin:2007ap, Sasaki:2008ha}.

In the SU(3)-flavor~[SU(3)$_F$] symmetric limit, one can classify the axial charges of baryons in terms of the constants of two types of couplings, $F$ and $D$~\cite{deSwart:1963gc}, as follows:
\begin{align}
\begin{split}
		&g_{A,NN}=F+D,\quad g_{A,\Xi\Xi}=D-F, \quad g_{A,\Sigma\Sigma}= 2F,\\
		&g_{A,\Lambda \Xi}=3F-D,\quad g_{A,\Sigma\Xi}=-(F+D),\\
		&g_{A,\Lambda N}=3F+D,\quad g_{A,\Sigma N}=D-F,\quad g_{A,\Lambda \Sigma}=2D.\\
\label{su3rel}
\end{split}
\end{align}%
This systematic classification, which phenomenologically works rather well but is not known {\it a priori} to hold, is expected to govern all the axial charges however as we move from the symmetric case to the realistic one, the SU(3)$_F$ breaking occurs as a result of the $s$-quark mass. The broken symmetry no longer provides a pattern for the couplings, and therefore they should be individually calculated based on the underlying theory, QCD.

Recently we have extracted the $\pi N\!N$, $\pi\Sigma\Sigma$, $\pi\Lambda\Sigma$, $K\Lambda N$ and $K \Sigma N $ coupling constants and the corresponding monopole masses in lattice QCD with two flavors of dynamical quarks~\cite{Erkol:2008yj}. We have found that the SU(3)$_F$ parameters have weak quark-mass dependence and thus the SU(3)$_F$ symmetry is broken by only a few percent. Our aim in this work is two-fold: We first concentrate on the coupling constants $\pi\Xi\Xi$, $K\Lambda\Xi$ and $K\Sigma\Xi$ in order to complete our program of calculating the pseudoscalar-meson--octet-baryon coupling constants from lattice QCD that we started in Ref.~\cite{Erkol:2008yj}. In the second part we evaluate the strangeness-conserving $N N$, $\Sigma\Sigma$, $\Xi\Xi$, $\Lambda\Sigma$ and the strangeness-changing $\Lambda N$, $\Sigma N $, $\Lambda\Xi$, $\Sigma\Xi$ axial charges in lattice QCD with two flavors of dynamical quarks. The evaluation of the coupling constants and the axial charges allows us to check whether the SU(3)$_F$ relations are well respected in the degenerate quark-mass case and to what extent this symmetry is broken as we restore the physical masses of quarks. We assume exact flavor-SU(2) symmetry and take $u$ and the $d$ quarks degenerate.

\section{The formulation and the lattice simulations}\label{sec2}
We refer the reader to Ref.~\cite{Erkol:2008yj} for the lattice formulation and the details of the calculations of pseudoscalar-meson--octet-baryon coupling constants. As for the axial charges, we consider the baryon matrix elements of the isovector axial-vector current $A_\mu=\overline{u}\gamma_\mu\gamma_5 u-\overline{d}\gamma_\mu\gamma_5 d$, which can be written in the form
\begin{align}\label{matel}
	\begin{split}
	\langle {\cal B}(p)|A_\mu|{\cal B}^\prime(p^\prime)\rangle=&C_{{\cal BB}^\prime} \bar{u}(p) \left[\gamma_\mu\gamma_5 G_{A,{\cal BB}^\prime}(q^2)\right.\\
	&\left.+\gamma_5\frac{q_\mu}{m_{\cal B}+m_{{\cal B}^\prime}} G_{P,{\cal BB}^\prime}(q^2) \right]u(p)
	\end{split}
\end{align}
where $q_\mu=p_\mu^\prime-p_\mu$ is the transferred four-momentum and $u(p)$ denotes the Dirac spinor for the baryon with four-momentum $p$ and $m_{\cal B}$. $G_{A,{\cal BB}^\prime}(p^2)$ and $G_{P,{\cal BB}^\prime}(p^2)$ are the baryon axial and induced pseudoscalar form factors, respectively. The isospin factors $C_{{\cal BB}^\prime}$ are given as $C_{NN}\equiv C_{\Xi\Xi}\equiv C_{N\Sigma}\equiv C_{\Sigma\Xi}=1$, $C_{\Sigma\Sigma}=-1/\sqrt{2}$, $C_{\Lambda\Sigma}\equiv C_{\Lambda\Xi} \equiv -C_{N\Lambda}=1/\sqrt{6}$.

The baryon axial charges are defined as the axial form factors at zero-momentum transfer, {\it viz.} $g_{A,{\cal BB}^\prime}=G_{A,{\cal BB}^\prime}(0)$. We compute the matrix element in Eq.~\eqref{matel} using the ratio
\begin{align}
\begin{split}\label{ratio}
	&R(t_2,t_1;{\bf p}^\prime,{\bf p};\Gamma;\mu)=\\
	&\quad\frac{\langle F^{{\cal B} {\cal A}_\mu {\cal B}^\prime}(t_2,t_1; {\bf p}^\prime, {\bf p};\Gamma)\rangle}{\langle F^{{\cal B}^\prime}(t_2; {\bf p}^\prime;\Gamma_4)\rangle} \left[\frac{\langle F^{{\cal B}}(t_2-t_1; {\bf p};\Gamma_4)\rangle}{\langle F^{{\cal B}^\prime}(t_2-t_1; {\bf p}^\prime;\Gamma_4)\rangle}\right.\\ 
	&\quad\left.\times\frac{\langle F^{{\cal B}^\prime}(t_1; {\bf p}^\prime;\Gamma_4)\rangle \langle F^{{\cal B}^\prime}(t_2; {\bf p}^\prime;\Gamma_4)\rangle}{\langle F^{{\cal B}}(t_1; {\bf p};\Gamma_4)\rangle \langle F^{{\cal B}}(t_2; {\bf p};\Gamma_4)\rangle} \right]^{1/2},
\end{split}
\end{align}
where the baryonic two- and three-point correlation functions are respectively defined as
\allowdisplaybreaks{
\begin{align}
	\begin{split}\label{twopcf}
	&\langle F^{{\cal B}}(t; {\bf p};\Gamma_4)\rangle=\sum_{\bf x}e^{-i{\bf p}\cdot {\bf x}}\Gamma_4^{\alpha\alpha^\prime} \\
	&\qquad\times \langle \text{vac} | T [\eta_{\cal B}^\alpha(x) \bar{\eta}_{{\cal B}^\prime}^{\alpha^\prime}(0)] | \text{vac}\rangle,
	\end{split}\\
	\begin{split}
	&\langle F^{{\cal B A_\mu B^\prime}}(t_2,t_1; {\bf p}^\prime, {\bf p};\Gamma)\rangle=-i\sum_{{\bf x_2},{\bf x_1}} e^{-i{\bf p}\cdot {\bf x_2}} e^{i{\bf q}\cdot {\bf x_1}} \\
	&\qquad\times \Gamma^{\alpha\alpha^\prime} \langle \text{vac} | T [\eta_{\cal B}^\alpha(x_2) A_\mu(x_1) \bar{\eta}_{{\cal B}^\prime}^{\alpha^\prime}(0)] | \text{vac}\rangle,
	\end{split}
\end{align}
}%
with $\Gamma\equiv\gamma_3 \gamma_5 \Gamma_4$ and $\Gamma_4\equiv (1+\gamma_4)/2$. The baryon interpolating fields are given as
\begin{align}
	\begin{split}\raisetag{60pt}
		\eta_N(x)&=\epsilon^{abc}[u^{T a}(x) C \gamma_5 d^b(x)]u^c(x),\\
		\eta_\Xi(x)&=\epsilon^{abc}[s^{T a}(x) C \gamma_5 d^b(x)]s^c(x),\\
		\eta_\Sigma(x)&=\epsilon^{abc}[s^{T a}(x) C \gamma_5 u^b(x)]u^c(x),\\
		\eta_\Lambda(x)&=\frac{1}{\sqrt{6}}\epsilon^{abc}\{[u^{T a}(x) C \gamma_5 s^b(x)]d^c(x)-[d^{T a}(x)C\\
		&\quad \times \gamma_5 s^b(x)]u^c(x)+2[u^{T a}(x) C \gamma_5 d^b(x)]s^c(x)\},
	\end{split}
\end{align}
where $C=\gamma_4\gamma_2$ and $a$, $b$, $c$ are the color indices. $t_1$ is the time when the meson interacts with a quark and $t_2$ is the time when the final baryon state is annihilated. The ratio in Eq.~(\ref{ratio}) reduces to the desired form when $t_2-t_1$ and $t_1\gg a$, {\it viz.}
\begin{equation}\label{desratio}
	R(t_2,t_1;{\bf 0},{\bf p};\Gamma;\mu)\xrightarrow[t_2-t_1\gg a]{t_1\gg a} \sqrt{\frac{E+m}{2m}}\,G_{A,{\cal BB}^\prime}(Q^2),
\end{equation}
where $m$ and $E$ are the mass and the energy of the initial baryon and $Q^2=-q^2$. We apply a procedure of seeking plateau regions as a function of $t_1$ in the ratio \eqref{desratio} and calculating the axial form factors $G_{A,{\cal BB}^\prime}(Q^2)$ at $Q^2=0$ in order to extract the axial charges $g_{A,{\cal BB}^\prime}$.

We employ the same lattice configuration as in our previous work in Ref~\cite{Erkol:2008yj}. It is a $16^3\times 32$ lattice with two flavors of dynamical quarks and the gauge configurations we use have been generated by the CP-PACS collaboration~\cite{AliKhan:2001tx} with the renormalization group improved gauge action and the mean-field improved clover quark action. We use the gauge configurations at $\beta=1.95$ with the clover coefficient $c_{SW}=1.530$, which give a lattice spacing of $a=0.1555(17)$ fm ($a^{-1}=1.267$ GeV) as determined from the $\rho$-meson mass. The simulations are carried out with four different hopping parameters for the sea and the $u$,$d$ valence quarks, $\kappa_{sea},\kappa_{val}^{u,d}=$ 0.1375, 0.1390, 0.1400 and 0.1410, which correspond to quark masses of $\sim$ 150, 100, 65, and 35~MeV, and we use 490, 680, 680 and 490 such gauge configurations, respectively. The hopping parameter for the $s$ valence quark is fixed to $\kappa_{val}^{s}=0.1393$ so that the Kaon mass is reproduced~\cite{AliKhan:2001tx}, which corresponds to a quark mass of $\sim 90$~MeV. We employ smeared source and smeared sink, which are separated by 8 lattice units in the temporal direction. Source and sink operators are smeared in a gauge-invariant manner with the root mean square radius of 0.6 fm. All the statistical errors are estimated via the jackknife analysis. The renormalization factors relevant to the axial currents are all computed in a perturbative manner: $Z_A=$0.2576, 0.2491, 0.2434 and 0.2377 at $\kappa=$ 0.1375, 0.1390, 0.1400 and 0.1410, respectively~\cite{AliKhan:2001tx}. 

Here we mention the systematic errors that could enter our results. Possible systematic errors arise from (1) the finite volume, (2) the perturbative estimation of renormalization factors, (3) the finite lattice spacing, (4) the wrong number of dynamical quarks, and (5) the unrealistic heavy quarks. 

For the finite volume effect, the present spatial lattice extent is 16 (about 2.5 fm) and the pion mass ranges from 0.440 to 0.899 in lattice unit (from 550 MeV to 1.15 GeV), which gives $7\leq m_\pi L \leq 14$. Serious finite volume artifact for the nucleon axial coupling seems to appear only when $m_\pi L \leq 7$~\cite{Yamazaki:2009zq}; so we expect small finite volume effects in our present calculations. 

We estimate the renormalization factors in a perturbative way, which gives rise to {\cal O}(10)\% errors in the nucleon vector charge or the pseudoscalar meson decay constants~\cite{Takahashi:2008fy}. In order to reduce such systematic errors, we evaluate the ratios of the couplings, which would be less dependent on these factors. For consistency check, we compare our results with the those of Lin and Orginos~\cite{Lin:2007ap} and find very good agreement (See the discussion below). Then, the systematic errors from (1)-(4) are considered to be small in comparison with statistical errors. 

Although our (light) quarks are heavier than real u,d-quarks, we are mainly focused on SU(3)$_F$ breaking pattern. Our present setup covers a wide range of quark masses including the SU(3)$_F$ symmetric point ($m_u = m_d = m_s$). On the other hand, it is difficult to estimate the breaking pattern at the chiral limit. The physics at the physical point could be accessed with much lighter quarks as well as more realistic lattice setups, which is planned for a future work.

\section{Results and Discussion}\label{sec3}
%%%%%%%%% table of results %%%%%%%%%%%%%%%%%%%%%%
\begin{table*}[ht]
	\caption{The fitted values of the $\pi\Xi\Xi$, $K\Lambda\Xi$ and $K\Sigma\Xi$ coupling constants and the corresponding monopole masses normalized with $g_{\pi N\!N}$ and $\Lambda_{\pi N\!N}$, respectively. Here, we define $g^R_{M {\cal BB}^\prime}=g_{M {\cal BB}^\prime}/g_{\pi N\!N}$ and $\Lambda^R_{M {\cal BB}^\prime}=\Lambda_{M {\cal BB}^\prime}/\Lambda_{\pi N\!N}$. 
}
%	\addtolength{\tabcolsep}{6pt}
\begin{center}
\begin{tabular*}{1.0\textwidth}{@{\extracolsep{\fill}}ccccccccccc}
		\hline\hline 
		$\kappa^{u,d}_{val}$ & $g^R_{\pi\Xi\Xi}$ & $g^R_{K\Lambda\Xi}$ & $g^R_{K\Sigma\Xi}$ & $\Lambda^R_{\pi\Xi\Xi}$ & $\Lambda^R_{K\Lambda \Xi}$ & $\Lambda^R_{K\Sigma \Xi}$ \\[0.5ex]
		\hline 
		0.1375 &  -0.227(18) & 0.334(15)& -1.025(20) & 0.687(130)& 1.030(164)& 0.884(39)\\
		0.1390 & -0.216(14)& 0.348(16)& -1.037(18)& 0.889(206)& 0.908(149)& 0.891(44)\\
		0.1393 & -0.217(14)& 0.347(16)& -1.036(19)& 0.882(208)& 0.918(157)& 0.891(49)\\
		0.1400 & -0.245(13)& 0.313(14)& -0.998(10)& 0.825(111)& 1.085(146)& 1.044(28)\\
		0.1410 & -0.273(26)& 0.291(25)& -0.963(48)& 0.896(148)& 1.186(242)& 1.237(99)\\
		\hline\hline
\end{tabular*}
	\label{res_table}
\end{center}
\end{table*}
%%%%%%%%%%%%%%%%%%%%%%%%%%%%%%%%%%%%%%%%%%%%%%%%%%%%%%%%%%%%%%%%%%
%%%%%%%%%%%%%%%%%%%%%% Figure 1  %%%%%%%%%%%%%%%%%%%%%%
\begin{figure}[th]
	\includegraphics[width=0.5\textwidth,height=0.6\textheight]{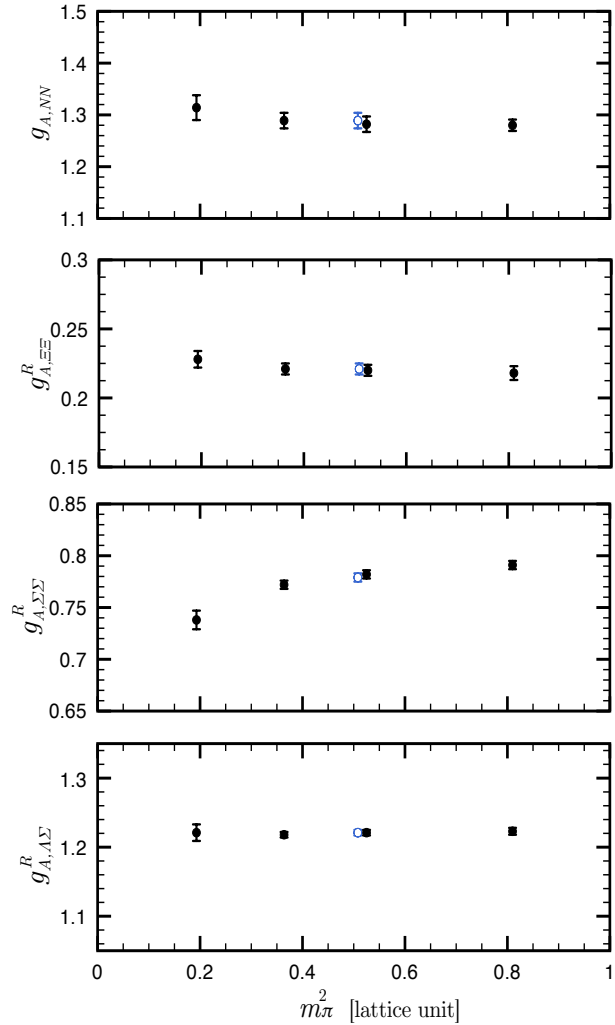}
	\caption{\label{sc} The $N\!N$ axial charge together with $\Xi\Xi$, $\Sigma\Sigma$, $\Lambda \Sigma$ axial charges normalized with $g_{A, N\!N}$ as a function of $m_\pi^2$. The empty circle denotes the SU(3)$_F$ limit.}
\end{figure}	
%%%%%%%%%%%%%%%%%%%%%%%%%%%%%%%%%%%%%%%%%%%%%%%%%%%%%%%%%%%%%%%%%%
%%%%%%%%%%%%%%%%%%%%%% Figure 2  %%%%%%%%%%%%%%%%%%%%%%
\begin{figure}[th]
	\includegraphics[width=0.5\textwidth,height=0.6\textheight]{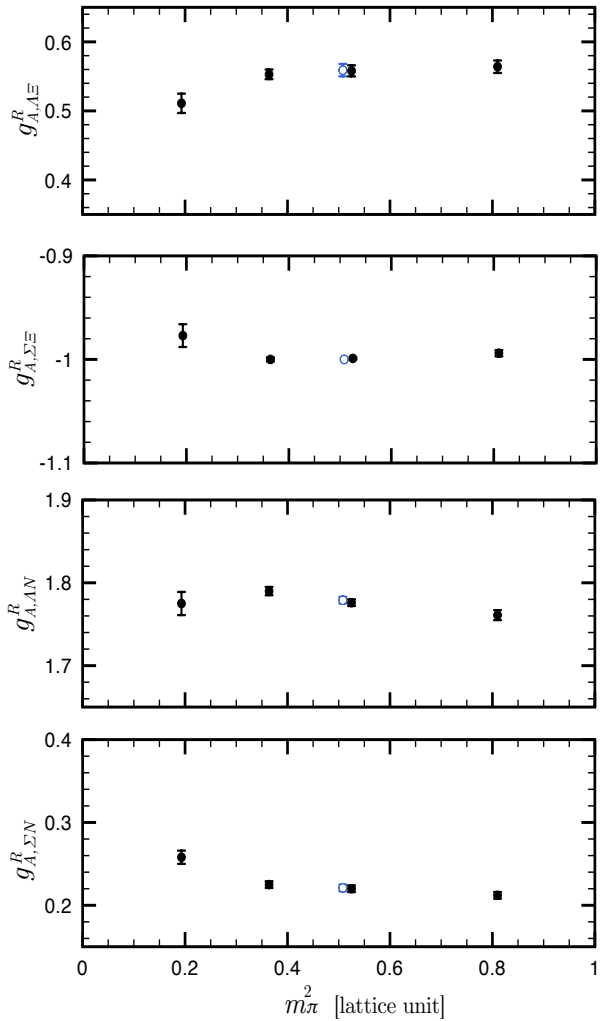}
	\caption{\label{sch} Same as Fig.~\ref{sc} but for strangeness-changing $\Lambda\Xi$, $\Sigma\Xi$, $\Lambda N$ and $\Sigma N $ axial charges.}
\end{figure}	
%%%%%%%%%%%%%%%%%%%%%%%%%%%%%%%%%%%%%%%%%%%%%%%%%%%%%%%%%%%%%%%%%%
We first concentrate on the coupling constants $g_{\pi\Xi\Xi}$, $g_{K\Lambda\Xi}$ and $g_{K\Sigma\Xi}$. The SU(3)$_F$ relations for these pseudoscalar coupling constants are given as 
\begin{equation}\label{su3ps}
	g_{\pi\Xi\Xi}=-g(1-2\alpha),~g_{K\Lambda\Xi}=\frac{1}{\sqrt{3}}g(4\alpha-1),~ g_{K\Sigma\Xi}=-g,
\end{equation}
where $g\equiv g_{\pi NN}$ and $\alpha$ is the $F/(F+D)$ ratio of the pseudoscalar octet. We extract these coupling constants, $g_{M {\cal BB}^\prime}$, and the corresponding monopole masses, $\Lambda_{M {\cal BB}^\prime}$, for each $\kappa_{val}^{u,d}$. Our results are presented in Table~\ref{res_table}: We give the fitted values of the $\pi\Xi\Xi$, $K\Lambda\Xi$, and $K\Sigma \Xi$ coupling constants and the corresponding monopole masses normalized with $g_{\pi N\!N}$ and $\Lambda_{\pi N\!N}$, respectively. In Table~\ref{res_table}, $g^R_{M {\cal BB}^\prime}$ and $\Lambda^R_{M {\cal BB}^\prime}$ denote $g_{M {\cal BB}^\prime}/g_{\pi N\!N}$ and $\Lambda_{M {\cal BB}^\prime}/\Lambda_{\pi N\!N}$, respectively. As in Ref.~\cite{Erkol:2008yj} we expect that the systematic errors cancel out to some degree in the ratios of the coupling constants and those of the monopole masses.

In the SU(3)$_F$ limit, where $\kappa_{val}^{u,d}\equiv\kappa_{val}^s = 0.1393$, the SU(3)$_F$ relations in Eq.(\ref{su3ps}) are exact and all the coupling constants are well reproduced with $\alpha=0.395(6)$, which is obtained by a global fit including the coupling constants obtained in Ref.~\cite{Erkol:2008yj}. In the SU(3)$_F$ broken case, we observe that our conclusion in Ref.~\cite{Erkol:2008yj} for the pseudoscalar-meson--baryon coupling constants holds as well for the coupling constants in question here: The quark-mass dependences for $g^R_{M{\cal BB'}}$ and $\Lambda^R_{M{\cal BB'}}$ are not large and the ratios of the coupling constants, $g^R_{M{\cal BB'}}$, are similar in value to those in the SU(3)$_F$ symmetric limit, and the monopole-mass ratios, $\Lambda^R_{M{\cal BB'}}$, are almost unity independently of the quark masses. This confirms that the SU(3)$_F$ breaking is small for pseudoscalar-meson--baryon coupling constants at the quark masses we consider.

We next concentrate on the strangeness-conserving and strangeness-changing axial charges of baryons. The axial charge of nucleon in the present setup is about 10\% overestimated~\cite{Edwards:2005ym}, which would be due to the perturbative estimation of renormalization factors~\cite{AliKhan:2001tx}. The present lattice spacing is about 0.15 fm, which is far from the continuum limit. In fact, the vector charge of nucleon as well as the decay constants obtained with the same setup as ours show {\cal O}(10)\% deviation from physical values~\cite{AliKhan:2001tx,Takahashi:2008fy}. Hence, we evaluate the ratios of axial charges (charges normalized with the axial charge of nucleon) rather than the bare values, so that we expect the cut-off artifacts in the renormalization factors to cancel to some extent. 

%%%%%%%%%%%%%%%%%%%%%% Figure 3  %%%%%%%%%%%%%%%%%%%%%%
\begin{figure}[th]
	\includegraphics[width=0.5\textwidth]{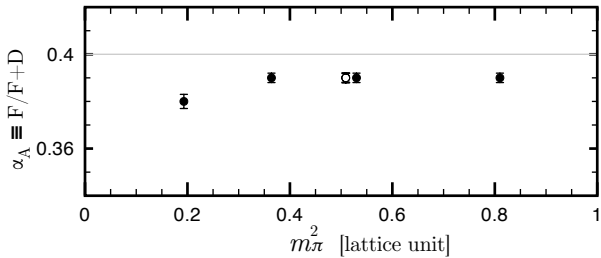}
	\caption{\label{alpha} $\alpha_A=F/F+D$ ratio as a function of $m_\pi^2$. The empty circle denotes the SU(3)$_F$ limit.}
\end{figure}	
%%%%%%%%%%%%%%%%%%%%%%%%%%%%%%%%%%%%%%%%%%%%%%%%%%%%%%%%%%%%%%%%%%

In Table~\ref{res_gA_table}, we give the fitted values of the $N\!N$ axial charge, $g_{A, N\!N}$, together with the fitted values of the strangeness-conserving $\Xi\Xi$, $\Sigma\Sigma$, $\Lambda \Sigma$ and strangeness-changing $\Lambda\Xi$, $\Sigma\Xi$, $\Lambda N$ and $\Sigma N $ axial charges normalized with $g_{A, N\!N}$ for various quark masses and illustrate these in Figs.~\ref{sc} and~\ref{sch}. Similarly to the pseudoscalar-meson--baryon coupling constants in our previous analysis~\cite{Erkol:2008yj}, we expect that the systematic errors cancel out to some degree in the ratios of the axial charges. Here, we define $g^R_{A,{\cal BB}^\prime}=g_{A,{\cal BB}^\prime}/g_{A,N\!N}$. We also present the values of the ratios of the coupling constants $\alpha_A=F/F+D$ as obtained from a global fit. In the SU(3)$_F$ limit, where $\kappa_{val}^{u,d}\equiv\kappa_{val}^s = 0.1393$, we obtain $\alpha_A=F/F+D=0.390(2)$. The value of $\alpha_A$ has a weak quark-mass dependence and as we approach the chiral point $\alpha_A$ tends to decrease. We illustrate this behavior in Fig.~\ref{alpha}.

%%%%%%%%%%%%%%%%%%%%%% Figure 4  %%%%%%%%%%%%%%%%%%%%%%
\begin{figure}[th]
	\includegraphics[width=0.5\textwidth]{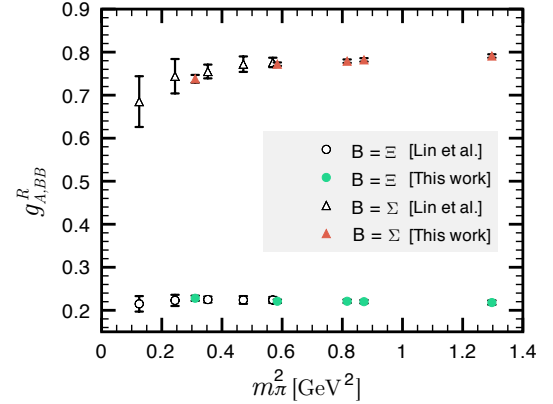}
	\caption{\label{comp} Comparison of our results for $g^R_{A,{ \Sigma\Sigma}}$ and $g^R_{A,{\Xi\Xi}}$ with those in Ref.~\cite{Lin:2007ap}}
\end{figure}	
%%%%%%%%%%%%%%%%%%%%%%%%%%%%%%%%%%%%%%%%%%%%%%%%%%%%%%%%%%%%%%%%%%

In Fig.~\ref{comp}, we compare our results for $g^R_{A,{ \Sigma\Sigma}}$ and $g^R_{A,{\Xi\Xi}}$ with those obtained from a lattice setup using staggered fermion action for the sea quarks and domain-wall fermions for the valence quarks~\cite{Lin:2007ap}. We observe the results from two different setups are in good agreement with each other.

%%%%%%%%%%%%%%%%%%%%%% Figure 5 %%%%%%%%%%%%%%%%%%%%%%
\begin{figure}[th]
	\includegraphics[width=0.5\textwidth]{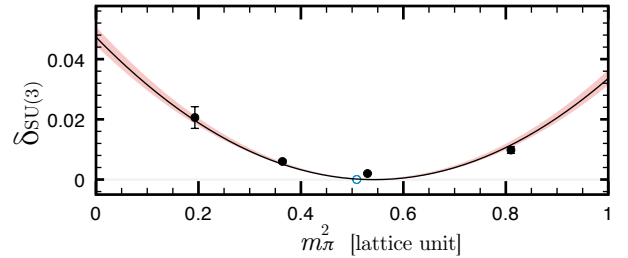}
	\caption{\label{SU3br} The SU(3)-breaking parameter, $\delta_{\text{SU}(3)}$, as a function of $m_\pi^2$. The empty circle denotes the SU(3)$_F$ limit.}
\end{figure}	
%%%%%%%%%%%%%%%%%%%%%%%%%%%%%%%%%%%%%%%%%%%%%%%%%%%%%%%%%%%%%%%%%%

%%%%%%%%% table of results gA %%%%%%%%%%%%%%%%%%%%%%
\begin{table*}[ht]
	\caption{The fitted value of the $N\!N$ axial charge together with the fitted values of the strangeness-conserving $\Xi\Xi$, $\Sigma\Sigma$, $\Lambda \Sigma$ and strangeness-changing $\Lambda\Xi$, $\Sigma\Xi$, $\Lambda N$ and $\Sigma N $ axial charges normalized with $g_{A, N\!N}$. Here, we define $g^R_{A,{\cal BB}^\prime}=g_{A,{\cal BB}^\prime}/g_{A,N\!N}$. We also give the fitted value of $F/F+D$ at each quark mass.}
%	\addtolength{\tabcolsep}{6pt}
\begin{center}
\begin{tabular*}{1.0\textwidth}{@{\extracolsep{\fill}}cccccccccc}
		\hline\hline 
		$\kappa^{u,d}_{val}$ & $g_{A,NN}$ & $g^R_{A,\Xi\Xi}$ & $g^R_{A,\Sigma\Sigma}$ &$g^R_{A,\Lambda \Sigma}$ & $g^R_{A,\Lambda\Xi}$& $g^R_{A,\Sigma\Xi}$ & $g^R_{A,\Lambda N}$ & $g^R_{A,\Sigma N}$ & $F/F+D$ \\[0.5ex]
		\hline 
		0.1375 & 1.284(11) & 0.218(05)& 0.791(04) & 1.223(05) & 0.564(09)& -0.994(03)& 1.761(06) & 0.212(04) &  0.390(2)\\
		0.1390 & 1.282(15) & 0.220(04)& 0.782(04) & 1.221(04) & 0.558(08)& -0.999(01)& 1.776(04)& 0.220(04) & 0.390(2)\\
		0.1393 & 1.280(15)& 0.221(04) & 0.779(04)& 1.221(04) & 0.559(09)& -1.000(01)& 1.779(04)& 0.221(04) & 0.390(2)\\
		0.1400 & 1.289(15)& 0.221(04)& 0.772(04) & 1.218(04) & 0.553(07)& -1.000(02)& 1.790(05)& 0.225(04) & 0.390(2)\\
		0.1410 & 1.314(24)& 0.228(06)& 0.738(09) & 1.221(12) & 0.511(14)& -0.977(11)& 1.775(14)& 0.258(08) & 0.380(3)\\
		\hline\hline
\end{tabular*}
	\label{res_gA_table}
\end{center}
\end{table*}
%%%%%%%%%%%%%%%%%%%%%%%%%%%%%%%%%%%%%%%%%%%%%%%%%%%%%%%%%%%%%%%%%%

In the SU(3)$_F$ broken case, the deviations in the coupling constants are not large and the values are similar in those at $\kappa=0.1393$. In order to quantify the SU(3)$_F$ breaking, we construct the following combinations:
\allowdisplaybreaks[1]{\begin{align}\label{su3br}
	& A_1=g^R_{A,\Xi\Xi}+g^R_{A,\Sigma\Sigma},\quad A_2=2 g^R_{A,\Xi\Xi}+g^R_{A,\Lambda\Xi},\\
	& A_3=(g^R_{A,\Xi\Xi}+g^R_{A,\Lambda N})/2,\quad A_4=g^R_{A,\Lambda\Sigma}-g^R_{A,\Xi\Xi},\notag\\
	& A_5=2 g^R_{A,\Sigma\Sigma}-g^R_{A,\Lambda\Xi},\quad A_6=g^R_{A,\Lambda N}-g^R_{A,\Sigma\Sigma},\notag\\
	& A_7=g^R_{A,\Sigma N}+g^R_{A,\Sigma\Sigma},\quad A_8= (g^R_{A,\Lambda\Sigma}+g^R_{A,\Sigma\Sigma})/2,\notag\\
	& A_9=(2 g^R_{A,\Lambda N}-g^R_{A,\Lambda\Xi})/3,\quad A_{10}= g^R_{A,\Lambda\Xi}+2g^R_{A,\Sigma N},\notag\\
	&A_{11}=(g^R_{A,\Lambda\Xi}+2 g^R_{A,\Lambda\Sigma})/3,\quad A_{12}= (g^R_{A,\Lambda N}+g^R_{A,\Sigma N})/2,\notag\\
	& A_{13}=(g^R_{A,\Lambda N}+g^R_{A,\Lambda\Sigma})/3,\quad A_{14}= g^R_{A,\Lambda\Sigma}-g^R_{A,\Sigma N},\notag\\
	& A_{15}=-g^R_{A,\Sigma\Xi}.\notag
\end{align}%
}
In the SU(3)$_F$ symmetric limit, the above equations satisfy $A_1\equiv A_2\equiv \ldots \equiv A_{15}=1$, which can be verified by inserting the coupling constants at $\kappa_{val}^{u,d}= 0.1393$ in Table~\ref{res_gA_table}. At other quark masses, the deviations from unity represent the amount of SU(3)$_F$ breaking. Inserting the values of the coupling constants corresponding to the lowest quark mass we consider in Table~\ref{res_table} into \eqref{su3br}, we find $A_1=0.966(07)$, $A_2=0.966(11)$, $A_3=1.002(07)$, $A_4=0.993(08)$, $A_5=0.965(10)$, $A_6=1.037(10)$, $A_7=0.996(05)$, $A_8=0.979(06)$, $A_9=1.013(07)$, $A_{10}=1.027(10)$, $A_{11}=0.984(08)$, $A_{12}=1.017(06)$, $A_{13}=0.999(06)$, $A_{14}=0.962(10)$, $A_{15}=0.977(10)$ which indicate a breaking in SU(3)$_F$ by less than 10\%, as we approach the chiral limit. Moreover, we define the average SU(3)$_F$ breaking as follows:
\begin{equation}
	\delta_{\text{SU}(3)}=\frac{1}{15} \sum_{n} \lvert 1-A_n \rvert,
\end{equation}
which amounts to $\delta_{\text{SU}(3)}=$0.010(1), 0.002(1), 0.006(1), and 0.021(4) for the quark masses at $\sim$ 150, 100, 65, and 35~MeV, respectively. This suggests for the axial charges of the octet baryons that SU(3)$_F$ is a good symmetry in the quark-mass range we consider, which is broken by only a few percent, similarly to the pseudoscalar-meson coupling constants. We have also tried a quadratic fit of $\delta_{\text{SU}(3)}$ and extracted $\delta_{\text{SU}(3)}=0.047(3)$ in the chiral limit. Fig.~\ref{SU3br} shows the value of $\delta_{\text{SU}(3)}$ as a function of $m_\pi^2$ and the chiral extrapolation with errors. As for $\alpha_A=F/F+D$, it seems to have a slightly smaller quark-mass dependence as compared to $\alpha=F/F+D$ of pseudoscalar-meson--baryon coupling constants~\cite{Erkol:2008yj}. It is interesting to note that $\alpha_A$ as extracted from axial charges is closer to 2/5, prediction from SU(6) quark model, in the present quark-mass range.

%%%%%%%%%%%%%%%%%%%%%% Figure 6 %%%%%%%%%%%%%%%%%%%%%%
\begin{figure}[th]
	\includegraphics[width=0.5\textwidth]{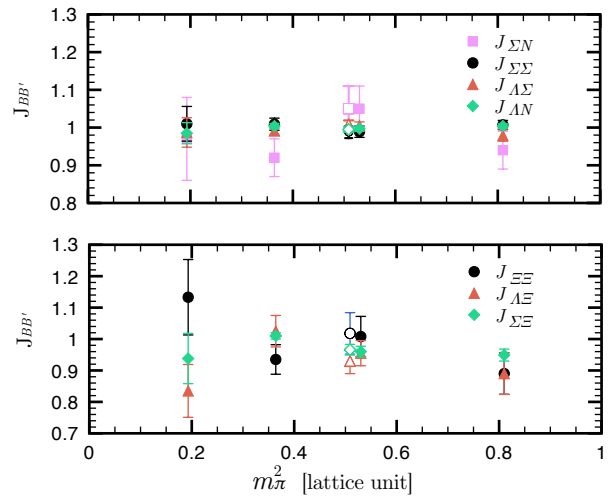}
	\caption{\label{gtr} $J_{\cal BB'}$ as defined in Eq.~\eqref{gtrform} as a function of $m_\pi^2$.}
\end{figure}	
%%%%%%%%%%%%%%%%%%%%%%%%%%%%%%%%%%%%%%%%%%%%%%%%%%%%%%%%%%%%%%%%%%

Assuming pion-pole dominance, the axial charges of octet baryons are related to their pseudoscalar-meson coupling constants via Goldberger-Treiman relations:
\begin{equation}\label{gtrel}
	f_M g_{M{\cal BB'}}={\cal S}_{\cal BB'} (m_{\cal B}+m_{\cal B'})/2\, g_{A,{\cal BB'}},
\end{equation}
where $f_M$ is the meson decay constant and ${\cal S}_{\cal BB'}$ are factors that fix our convention in choosing the isospin factors in Eq.\eqref{matel} with respect to Goldberger-Treiman relations: ${\cal S}_{NN}\equiv {\cal S}_{\Sigma\Sigma} \equiv {\cal S}_{\Sigma N}\equiv {\cal S}_{\Sigma\Xi}\equiv -{\cal S}_{\Xi\Xi}=1$, ${\cal S}_{\Lambda\Sigma}\equiv -{\cal S}_{\Lambda N} \equiv {\cal S}_{\Lambda\Xi}=1/\sqrt{3}$. We can check the validity of these relations by normalizing both sides with the Goldberger-Treiman relation of the nucleon and pion, \emph{viz.},
\begin{equation}
	f_\pi g_{\pi NN}=m_N\, g_{A,NN},
\end{equation}
and by constructing the following form:
\begin{equation}\label{gtrform}
	J_{\cal BB'}={\cal S_{BB'}}\frac{g^R_{A,{\cal BB'}}}{g^R_{M{\cal BB'}}}\frac{f_\pi}{f_{M}}\frac{(m_{\cal B}+m_{\cal B'})}{2m_N},
\end{equation}
in order to eliminate the systematic errors. If these relations are valid, then we should have $J_{\cal BB'}=1$. In Fig.~\ref{gtr}, we present $J_{\cal BB'}$ for all the couplings in question as a function of $m_\pi^2$. Since we have large error bars, it is difficult to reach a definite conclusion about the validity of the Goldberger-Treiman relations, however, we observe that these relations are rather good in the SU(3)$_F$-symmetric limit and discrepancies arise, in particular for $\Xi$ couplings, as we approach the chiral limit.

\section{Conclusion}\label{sec4}
We have evaluated the strangeness-conserving $N N$, $\Sigma\Sigma$, $\Xi\Xi$, $\Lambda\Sigma$ and the strangeness-changing $\Lambda N$, $\Sigma N $, $\Lambda\Xi$, $\Sigma\Xi$ axial charges in lattice QCD with two flavors of dynamical quarks and extended our earlier work in Ref.~\cite{Erkol:2008yj} so as to include the $\pi\Xi\Xi$, $K\Lambda\Xi$ and $K\Sigma\Xi$ coupling constants. We have extracted the ratios of the axial charges, which are supposed to be less prone to systematic errors. The ratios of the axial charges show rather weak quark-mass dependence. We have allowed an SU(3)$_F$ breaking by varying the quark masses. Our results suggest that the SU(3)$_F$ for axial charges is a good symmetry, which is broken by only a few percent. This conclusion is in agreement with what we have found for the pseudoscalar-meson couplings of the octet baryons. While we think that the present work reveals the SU(3)$_F$ pattern of axial-vector couplings of octet baryons, there are a number of improvements to be considered in a future work. Our lattice is still coarse by modern standards and quark masses are too large to reach a definite conclusion about SU(3)$_F$ breaking. Simulations with more realistic setups with smaller lattice spacing and larger lattice size employing much lighter quarks and a dynamical $s$-quark are under way~\cite{Aoki:2008sm}. It is an intriguing issue to see whether our findings in this work are retained in more realistic 2+1-flavor calculations with much lighter quark masses.

\acknowledgments
All the numerical calculations were performed on NEC SX-8R at CMC (Osaka university), SX-8 at YITP (Kyoto University), BlueGene/L (KEK), TSUBAME (TITech) and on National Center for High Performance Computing of Turkey (Istanbul Technical University). The unquenched gauge configurations employed in our analysis were all generated by CP-PACS collaboration~\cite{AliKhan:2001tx}. This work was supported in part by the Yukawa International Program for Quark-Hadron Sciences (YIPQS) and by KAKENHI (17070002, 19540275, 20028006 and 21740181)

%\bibliography{/Users/erkol/academy/lattice_paper/latticegA/mbcl}

\end{document}